\documentclass[10pt,final,doublecolumn]{IEEEtran}
\usepackage{amsmath}
\usepackage{latexsym}
\usepackage{graphicx}
\usepackage{bbding}
\usepackage{indentfirst}
\usepackage{algorithm,algorithmic}
\usepackage{setspace}
\usepackage{float}
\usepackage{epstopdf}
\usepackage{amssymb}
\usepackage{amsfonts}
\usepackage{enumerate}
\usepackage{multicol}
\usepackage{color}
\usepackage{slashbox}
\usepackage{bm}
\usepackage{amssymb}
\usepackage{stfloats}

\IEEEoverridecommandlockouts
\allowdisplaybreaks[4]

\begin{document}

\title{Feature-Aided Adaptive-Tuning Deep Learning for Massive Device Detection}
\author{
Xiaodan Shao, Xiaoming Chen, Yiyang Qiang, Caijun Zhong, and Zhaoyang Zhang
\thanks{Xiaodan Shao ({\tt shaoxiaodan@zju.edu.cn}), Xiaoming Chen ({\tt chen\_xiaoming@zju.edu.cn}), Yiyang Qiang ({\tt yiyang\_qiang@zju.edu.cn}), Caijun Zhong ({\tt caijunzhong@zju.edu.cn}), and Zhaoyang Zhang ({\tt ning\_ming@zju.edu.cn}) are with the College of Information Science and Electronic Engineering, Zhejiang University, Hangzhou 310027, China. }}
\maketitle
\begin{abstract}
With the increasing development of Internet of Things (IoT), the upcoming sixth-generation (6G) wireless network is required to support grant-free random access of a massive number of sporadic traffic devices. In particular, at the beginning of each time slot, the base station (BS) performs joint activity detection and channel estimation (JADCE) based on the received pilot sequences sent from active devices. Due to the deployment of a large-scale antenna array and the existence of a massive number of IoT devices, conventional JADCE approaches usually have high computational complexity and need long pilot sequences. To solve these challenges, this paper proposes a novel deep learning framework for JADCE in 6G wireless networks, which contains a dimension reduction module, a deep learning network module, an active device detection module, and a channel estimation module. Then, prior-feature learning followed by an adaptive-tuning strategy is proposed, where an inner network composed of the Expectation-maximization (EM) and back-propagation is introduced to jointly tune the precision and learn the distribution parameters of the device state matrix. Finally, by designing the inner layer-by-layer and outer layer-by-layer training method, a feature-aided adaptive-tuning deep learning network is built. Both theoretical analysis and simulation results confirm that the proposed deep learning framework has low computational complexity and needs short pilot sequences in practical scenarios.
\end{abstract}

\begin{IEEEkeywords}
6G, grant-free random access, active device detection, channel estimation, deep learning.
\end{IEEEkeywords}

\IEEEpeerreviewmaketitle
\section{Introduction}
Driven by the fast proliferation of Internet of Things (IoT), massive machine-type communication (mMTC) has been identified as one of the main use cases of the sixth-generation ($6$G) wireless networks \cite{iot}-\cite{shao1}. Generally speaking, mMTC has three fundamental characteristics, namely massive connectivity, sporadic data traffic and small payload \cite{Massiveaccess}. In the context of mMTC, conventional grant-based random access schemes lead to a high access latency and a prohibitive signaling overhead. To this end, grant-free random access schemes have been regarded as a candidate technology of $6$G wireless networks \cite{free, free1}. In particular, active devices transmit their data signals without a grant from the BS after sending pre-assigned pilot sequences. Hence, the key of grant-free random access is active device detection and channel estimation at the BS based on the received pilot sequences.

Due to the sporadic characteristic of IoT data traffic, active device detection is a typical sparse signal recovery problem. Specifically, active device detection is equivalent to the recovery of the sparse device state matrix from the noisy measurements. Therefore, many works attempt to address the problem of active device detection by using compressed sensing (CS) approaches \cite{CS}. For instance, the authors in \cite{ADC} proposed a generalized turbo signal recovery algorithm which was capable of achieving good performance and strong robustness in solving the active device detection problem with a mixed analog-to-digital converter (ADC) architecture. By exploiting both the active device sparsity and the chunk sparsity feature of the channel matrix, the authors in \cite{CRAN} proposed a modified Bayesian CS algorithm, which can further improve the detection performance. If channel information is available, the accuracy of active device detection can be significantly increased. Motivated by that, the authors in \cite{AMPO} and \cite{AMPliang1} proposed an approximate message passing (AMP) algorithm for active device detection by making use of the statistical information of wireless channels. It was proved that if the number of BS antennas was sufficiently large, the detection error asymptotically approached zero in the case that the elements of pilot sequences were independent and identically distributed (i.i.d.) Gaussian random variables with zero mean. To provide more pilot sequences with a given length, pilot sequences have to be non-i.i.d. For improving the estimation performance in the case of non-i.i.d. pilot sequences, a vector AMP (VAMP) algorithm was proposed in \cite{vampr} by introducing a linear minimum mean square error (LMMSE) estimator into AMP. Note that the aforementioned approaches in \cite{ADC}-\cite{vampr} performed active device detection based on the instantaneous received signals, which requires exceedingly long pilot sequences in the scenario of massive access. To tackle this problem, the covariance-based algorithms were proposed in \cite{cov}-\cite{cov3} to improve the performance of device activity detection. In specific, \cite{cov} formulated
device activity detection as an maximum likelihood (ML) estimation problem, in which the received signals at multiple antennas affected the detection results via their empirical covariance matrix. Then, \cite{cov1} proposed a joint device activity and data detection scheme based on the covariance matrix of the received signals, and analyzed the distribution of the estimation error in the massive MIMO regime. In fact, the covariance-based algorithm can outperform the AMP algorithm with the same length of pilot sequences. However, such a superior system performance comes from the expense of the use of a relatively large number of BS antennas. Although the algorithms in \cite{ADC}-\cite{cov2} can achieve good detection performance by exploiting the sparsity structure of the device state matrix, they impose high computational complexity due to large-dimensional matrix operators caused by the large-scale antenna array and the massive number of IoT devices \cite{shao0}. More importantly, these algorithms require long pilot sequences, which cannot satisfy the requirement of short-packet communications in the context of small IoT payload. As a result, active device detection in $6$G wireless networks has emerged as a challenging problem due to a large number of devices and the limited radio resources in wireless networks \cite{derrick, derrick1}.

To overcome these challenges, one can project the original detection problem in a high-dimensional space to a low-dimensional space by exploiting its specific structure \cite{shao}. In addition, if the apriori knowledge of the device state matrix that needs to be recovered is available in advance, the required length of the pilot sequences can be shortened for a given performance requirement. For example, the AMP algorithms improved the detection performance by modeling the device state matrix as a Bernoulli-Gaussian distributed random matrix based on the assumptions that uplink channels were Gaussian distributed \cite{AMPliang1}. These assumptions, however, still have limitations, since wireless channels in practical environments often exhibit much more complex statistical structures. Specifically, some research works in \cite{channel}-\cite{channel3} have shown that the channels in massive multiple-input multiple-output (MIMO) systems exhibit spatial sparsity, which can be approximated by the Gaussian mixture distribution. As a result, the interested device state matrix is a Bernoulli-Gaussian mixture distributed random matrix. Based on such a distribution, the detection performance can be improved in practical scenarios.  However, it is not trivial to obtain the parameters of the Bernoulli-Gaussian mixture distribution by employing traditional channel estimation approaches.

An effective approach for jointly obtaining and applying the distribution parameters is deep learning due to its powerful capabilities of data processing. Recently, deep learning has been widely adopted to design various advanced wireless communication techniques \cite{deep}-\cite{csl}. For instance, in \cite{oamp-net}, the authors proposed an orthogonal AMP-Net (OAMP-Net) algorithm which incorporated deep learning into the OAMP algorithm. The OAMP-Net algorithm can significantly improve the detection performance over Rayleigh and correlated MIMO channels.
The authors in \cite{mmlearn} proposed a millimeter-wave beam prediction scheme that combined machine learning tools and situational awareness to learn the beam information, including power and optimal beam index from past observations. A deep learning compressed sensing channel estimation scheme was proposed in \cite{csl}, where the channel estimation neural network was trained offline using simulated environments to predict the beamspace channel amplitude, then the channel was reconstructed based on the obtained indices of dominant beamspace channel entries. For active device detection, deep learning also has a great potential for performance enhancement. For example, the pilot matrix and the support recovery method were jointly designed in \cite{li} by using an auto-encoder deep learning network, where the auto-encoder comprised an encoder which mimicked the noisy linear measurement process and a decoder which approximately performed the sparse support recovery from the under-sampled linear measurements. Moreover, the authors in \cite{wei} proposed a deep learning algorithm to enforce the suspicious device to be inactive in each iteration of the AMP algorithm via employing the idea of list decoding in the field of error control coding. Unfortunately, the data-driven deep learning approaches in \cite{li} and \cite{wei} for active device detection require a large number of training data to achieve satisfactory performance. However, in the scenario of time-varying fading channels, the data-driven deep learning approaches are inapplicable as the channel coherence time is limited. In this context, this paper aims to design a model-driven deep learning framework for massive device detection. The main challenges lie in that how to exploit the specific feature, i.e., the parameters of complex distribution of the device state matrix in a low-dimensional space, how to reduce the performance loss caused by imperfect learning of the prior distribution parameters, and how to design a scheme to train the parameters of the detection algorithm. The contributions of this paper are as follows:
\begin{enumerate}
\item This paper proposes a novel deep learning framework for massive device detection in $6$G wireless networks, which contains a dimension reduction module, a deep learning network module, an active device detection module, and a channel estimation module. The proposed deep learning framework can perform joint activity detection and channel estimation with a finite number of training data.

\item This paper introduces an adaptive-tuning module in deep learning network by combing EM and back-propagation to adaptively tune the noise precision and learn the distribution parameters of the device state matrix. In addition, by exploiting prior features that the elements of device state matrix follow the Bernoulli-Gaussian mixture distribution, this paper derives a new denoiser for improving the detection performance, which is different from the original denoisers in the existing deep learning networks.

\item Integrating the prior-feature learning and adaptive-tuning module, this paper designs a feature-aided adaptive-tuning deep learning (FAT-DL) network including inner and outer networks to solve the massive device detection problem. By designing the layer-by-layer training method, a number of received data at the BS is used as the training data to obtain the feature, namely distribution parameters. Moreover, extensive simulation results confirm the effectiveness of the proposed algorithm in massive device detection problem.
\end{enumerate}

The rest of this paper is organized as follows: Section II gives a brief introduction of $6$G wireless networks with a massive number of sporadic traffic devices. Section III proposes a deep learning framework for massive device detection. Next, a feature-aided adaptive-tuning deep learning network is designed in Section IV. Extensive simulation results are shown in Section V. Finally, Section VI concludes the paper.

\emph{Notations}: We use bold letters to denote matrices or vectors, non-bold letters to denote scalars, $(\cdot)^T$ to denote transpose, $(\cdot)^H$ to denote conjugate transpose, $\mathbb{E}[\cdot]$ to denote expectation, $\mathrm{var}(\cdot)$ to denote the variance, $\|\cdot\|_2$ to denote the $l_2$-norm of a vector, $\|\cdot\|_F$ to denote the Frobenius norm of a matrix, $\mathbb{E}(\cdot |\cdot )$ to denote the conditional expectation operator, $\mathbb{C}^{A\times B}$ to denote the space of complex matrices of size $A\times B$, $|\cdot|$ to denote the absolute value. For a matrix $\mathbf{S}$, $\mathbf{s}_r$ denotes its $r$th column and $s_{nr}$ denotes its element in the $n$th row and $r$th column. $\cdot$ denotes multiplication.

\section{System Model and Problem Formulation}
\begin{figure}[t]
  \centering
\includegraphics [width=0.45\textwidth] {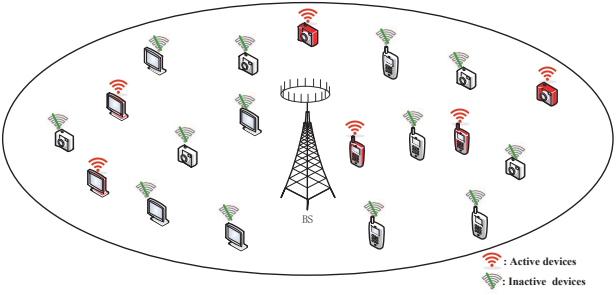}
\caption{A $6$G wireless network with sporadic traffic devices.}
\label{MUD}
\end{figure}
This paper considers a single-cell $6$G wireless network, where a BS equipped with $M$ antennas serves $N$ single-antenna IoT devices distributed in the network. In $6$G wireless networks, the density of IoT devices is usually very large, i.e., 10 per m$^2$. However, due to the burst characteristic of IoT applications, only a fraction of IoT devices are active at any given time slot, as shown in Fig. \ref{MUD}. In this context, a grant-free random access scheme is applied to jointly detect active devices and to estimate their corresponding channel state information (CSI). Specifically, at the beginning of each time slot, the active IoT devices simultaneously send predetermined pilot sequences to the BS, and then the BS performs JADCE based on the received signals. In a given time slot, $\mathcal{K}$ is used to denote the collection of active devices with $K=\left |\mathcal{K} \right |$ being the number of active devices. For convenience, define $\alpha_k$ as the activity indicator with ${\alpha _k} = 1$ if the $k$th device is active, and ${\alpha _k} = 0$ otherwise.

We adopt $\mathbf{h}_n \in \mathbb{C}^{M}$ to denote the channel vector from the $n$th device to the BS. It is assumed that the channels remain constant in a time slot and independently fade over time slots. A unique pilot sequence $\mathbf{a}_{n}=[a_{n,1},\cdots,a_{n,L}]^T\in \mathbb{C}^{L\times1}$ is assigned to the $n$th device, where $L$ is the pilot length. Let $\mathbf{E}\in \mathbb{C}^{L\times M}$ be the independent and identically distributed (i.i.d.) additive white Gaussian noise (AWGN) matrix, whose element $e_{l,m}$ follows the distribution $\mathcal{CN}(e_{l,m};0,{\sigma}^2)$, i.e., the complex Gaussian distribution with zero mean and variance $\sigma^2$. Besides, $\xi_n=Lp^p_n$ denotes the total pilot transmit energy with $p^p_n$ being the pilot transmit power of the $n$th device. Define $\mathbf{X}=[\mathbf{x}_1,...,\mathbf{x}_N]^T\in\mathbb{C}^{N\times M}$ with $\mathbf{x}_n={\alpha}_n\xi_n\mathbf{h}_n\in\mathbb{C}^{M\times1}$ as the device state matrix and $\mathbf{A}=[\mathbf{a}_1,...,\mathbf{a}_N]\in\mathbb{C}^{L\times N}$ as the pilot matrix. Thus, the received signal $\mathbf{Y}\in \mathbb{C}^{L\times M}$ at the BS can be cast as
\begin{eqnarray}
\label{eqinter5}
\mathbf{Y}
=\sum_{n=1}^{N}\alpha_n\xi_n\mathbf{a}_n\mathbf{h}_n^T+\mathbf{E}=\mathbf{AX+E}.
\end{eqnarray}
Based on the received signal $\mathbf{Y}$, the BS first performs active device detection, which is equivalent to recovering the sparse and low-rank device state matrix $\mathbf{X}$. Then, the corresponding CSI can be obtained after active device detection. Since pilot sequences are not orthogonal of each other, the recovery of $\mathbf{X}$ from the noisy measurement $\mathbf{Y}$ is not a trivial task. Especially in $6$G wireless networks, there are a large-scale antenna array at the BS and a massive number of IoT devices, resulting in high computational complexity in the recovery of $\mathbf{X}$ based on the conventional activity detection approaches.
In the following, we will propose a novel deep learning framework for recovering the device state matrix $\mathbf{X}$.

\section{A Novel Deep Learning Framework}
\begin{figure*}
  \centering
\includegraphics [width=1 \textwidth] {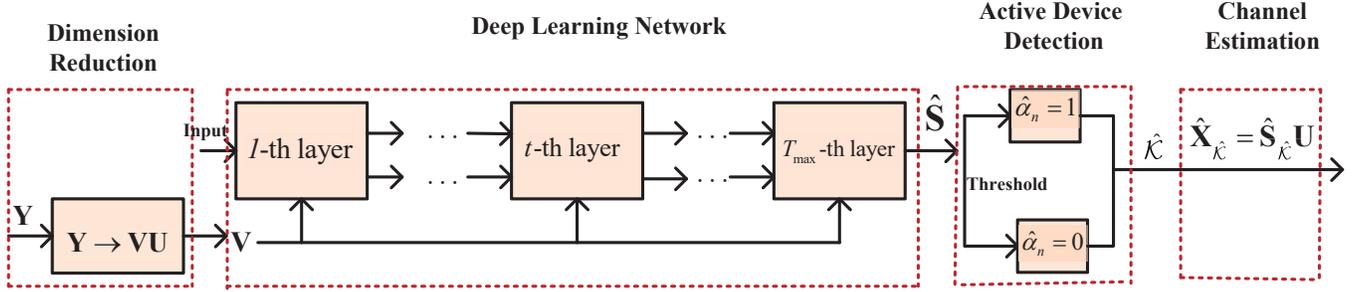}
\caption{The architecture of the proposed deep learning framework for massive access. Wherein, the deep learning network for the estimation of the transformed device state matrix consists of $T_{\max}$ layers, and each layer has the same structure.}
\label{architecture}
\end{figure*}

In this section, we design a deep learning framework for massive device detection in $6$G wireless networks. The designed framework aims to address two critical issues in conventional massive device detection approaches. Firstly, the conventional device detection approaches are only applicable to the scenarios of Rayleigh fading channels \cite{AMPliang1}, which limits their applicabilities significantly. Secondly, the conventional device detection approaches have high computational complexity.
As shown in Fig. \ref{architecture}, the proposed deep learning framework for massive device detection comprises four modules, i.e., a dimension reduction module, a deep learning network module, an active device detection module, and a channel estimation module. In what follows, we introduce these four modules.

\subsection{Dimension Reduction}
Since the BS of $6$G wireless networks is equipped with a large-scale antenna array, massive device detection based on $\mathbf{Y}$ imposes prohibitive computational complexity. Considering $\mathbf{X}\in\mathbb{C}^ {N \times M}$ is simultaneously sparse and low-rank, namely its rank $r^e = \text{rank}(\mathbf{X})\leq \min\{M,N\}$, we carry out dimension reduction before performing any sparse signal recovery. In particular, we can transform the original detection problem in (\ref{eqinter5}) to a low-dimensional space, where the unknown matrix is of size $N\times r^e$. Consequently, the solution of the original problem can be recovered based on that of the low-dimensional problem. Specifically, we conduct dimension reduction as follows \cite{shao}:

\begin{enumerate}
\item
The received signal $\mathbf{Y}$ in \eqref{eqinter5} is partitioned into a signal space and its null space by singular value decomposition (SVD), namely $\mathbf{Y}=\mathbf{S}_\mathrm{sd}\mathbf{V}_\mathrm{sd}\mathbf{D}_\mathrm{sd}^H$. Let $\mathbf{V}=\mathbf{S}_{r^e}\mathbf{V}_{r^e}$, where $\mathbf{S}_{r^e}$ is the first $r^e$ columns of $\mathbf{S}_\mathrm{sd}$, and $\mathbf{V}_{r^e}$ is a square matrix consisting of the first $r^e$ rows and the first $r^e$ columns of $\mathbf{V}_\mathrm{sd}$. Let $\mathbf{U}$ be the first $r^e$ rows of $\mathbf{D}_\mathrm{sd}^H$. Then, the signal space is constructed as
\begin{equation}\label{signalspace}
\mathbf{VU}=\mathbf{A}\mathbf{X}+\mathbf{E}_{X},
\end{equation}
where $\mathbf{E}_{X}$ is the noise incorporated in the signal space, $\mathbf{V}\in \mathbb{C}^{L\times r^e}$ with $\text{rank}(\mathbf{V}) = r^e$, and $\mathbf{U}\in \mathbb{C}^{r^e\times M}$ with $\mathbf{UU}^H=\mathbf{I}$.

\item
Based on \eqref{signalspace}, an equivalent form to the original input-output model (\ref{eqinter5}) in a low-dimensional space is constructed as
\begin{equation}\label{samd}
  \mathbf{V}=\mathbf{AS}+\mathbf{E}_{S},
\end{equation}
with $\mathbf{S}=\mathbf{X}\mathbf{U}^H$ and $\mathbf{E}_{S}=\mathbf{E}_{X}\mathbf{U}^H$. The transformed device state matrix $\mathbf{S}$ can be recovered uniquely in the low-dimensional space based on \eqref{samd}. Then, the originally concerned device state matrix $\mathbf{X}$ is obtained by letting $\mathbf{X}=\mathbf{S}\mathbf{U}$, where $\mathbf{S}$ has the same distribution parameters with $\mathbf{X}$.
\end{enumerate}

Applying the above dimension reduction, the transformed device state matrix $\mathbf{S}$ in the low-dimensional space can be adopted for active device detection. Since the number of BS antennas in $6$G wireless networks is huge, such a dimension reduction can decrease the computational complexity substantially during the recovery of the device state matrix. Importantly, $\mathbf{S}$ has the same distribution parameters with $\mathbf{X}$, a feature that will be exploited for improving detection performance in this paper. Different from this paper, the work in \cite{shao} mainly applied the full column rank property of transformed device state matrix $\mathbf{S}$ after the dimension reduction.
Note that the result in (\ref{signalspace}) is a high signal-to-noise ratio (SNR) characterization of the received signal. Such a high-SNR approximation is a reasonable assumption in $6$G wireless networks with massive connectivity due to the limited interference.

\subsection{Deep Learning Network}
After dimension reduction, a deep learning network module is utilized to estimate the transformed device state matrix $\mathbf{S}$. In general, the estimator is designed based on the distribution of the device state matrix. Even with dimension reduction, $\mathbf{S}$ has the same distribution parameters as $\mathbf{X}$. Commonly, $\mathbf{X}$ is assumed to be Bernoulli-Gaussian distributed if the channels experience Rayleigh fading \cite{AMPliang1}. However, in practical environments, the channels may have an irregular distribution. To enhance the robustness of the estimator, we model $\mathbf{X}$ as a Bernoulli-Gaussian mixture distributed random matrix \cite{bgm}. With the prior knowledge information, we design a feature-aided adaptive-tuning deep learning (FAT-DL) algorithm based on the architecture of the VAMP algorithm for the design of the estimator in Section IV.

\subsection{Active Device Detection}
Next, given the output of the deep learning network model $\hat{\mathbf{S}}$, we can detect the active devices based on a judgement threshold. In specific, we determine the estimated activity indicator $\hat{\alpha}_k$ as follows:
\begin{eqnarray}\label{thr}
\hat{\alpha}_k=\begin{cases}
1, & \text{ if } \left \| \hat{\mathbf{S}}(k,:) \right \|_2^2\geq v^2r^e \\
0, & \text{ if } \left \| \hat{\mathbf{S}}(k,:)\right \|_2^2< v^2r^e
\end{cases}
\end{eqnarray}
where $\hat{\mathbf{S}}(k,:)$ is the $k$th row of the estimation of device state matrix, i.e., $\hat{\mathbf{S}}$, $v^2r^e$ is the threshold, and $v=v_1\max(|\hat{s}_{nr}|), \forall n\in \{1,\cdots,N\},\forall r \in\{1,\cdots,r^e\}$ with $\max(|\hat{s}_{nr}|)$ being the maximum absolute value of element of $\hat{\mathbf{S}}$. Herein, $v_1$ is set according to the considered channel model. In particular, $v_1$ is set as the ratio of the minimum and the maximum amplitudes of the generated channel coefficients, which has been widely adopted in \cite{amp_gao} and \cite{mmwave}. The threshold in \eqref{thr} implies that if the energy of $\hat{\mathbf{S}}(k,:)$ is not less than the $r^e$ fold of the minimum element energy in statistical sense, the $k$th device is declared active.

\subsection{Channel Estimation}
Once the active devices are determined, the estimate of original interested device state matrix can be obtained by letting ${\hat{\mathbf{X}}}=\hat{\mathbf{S}}\mathbf{U}$. Correspondingly, the CSI of the $k$th active device $\hat{\mathbf{h}}_{k}$ can be estimated as
\begin{equation}
\label{gainesti}
\hat{\mathbf{h}}_{k}=\hat{\mathbf{x}}_{k}/\sqrt{\xi_k},
\end{equation}
where $\hat{\mathbf{x}}_{k}$ is the $k$th row of $\hat{\mathbf{X}}$.

Overall, the proposed deep learning framework combines the apriori knowledge of the device state matrix with a deep learning network to improve the robustness against a wider distributions range of the device state matrix and the pilot matrix, also conducts the dimension reduction operator to decrease the demand in computation resources and training time. In the following, we provide an effective algorithm to design the deep learning network module of the proposed framework for massive device detection in $6$G wireless networks.

\section{Feature-Aided Adaptive-Tuning Deep Learning Network Design}
In this section, we propose a feature-aided adaptive-tuning deep learning (FAT-DL) network to recover the transformed device state matrix by exploiting the prior feature of the device state matrix, and combining the AWGN precision adaptive-tuning and variable back-propagation to boost the learning performance. The VAMP algorithm is a powerful approach for solving the JADCE problem in high-dimensional settings \cite{vampr}, however, it can not adapt to general complex channel settings for JADCE problem. In what follows, we first briefly review the VAMP algorithm, and then design the FAT-DL network based on the architecture of VAMP.

\subsection{VAMP Algorithm}
Based on the low-dimensional received signal $\mathbf{V}$, $\mathbf{S}$ can be recovered by using the VAMP algorithm \cite{vampr}. In general, the VAMP algorithm performs activity detection by considering each column of the received signal in parallel and assuming that the distribution parameters of the device state matrix are known. The derivation of VAMP is through an approximation of a non-loopy graph with vector-valued nodes. For the low-dimensional received signal $\mathbf{V}$, the non-loopy graph is constructed based on the following factorization
\begin{equation}\label{xy}
 p(\mathbf{v}_r,\mathbf{s}_r)=p(\mathbf{s}_r)\mathcal{CN}(\mathbf{v}_r;\mathbf{As}_r,{\sigma}^2\mathbf{I}),
\end{equation}
where $\mathbf{v}_r$ and $\mathbf{s}_r$ are the $r$th column of $\mathbf{V}$ and $\mathbf{S}$, respectively. $p(\mathbf{s}_r)$ is the probability density function of $\mathbf{s}_r$. $\mathcal{CN}(\mathbf{v}_r;\mathbf{As}_r,{\sigma}^2\mathbf{I})$ represents the probability density function of the complex Gaussian distributed random variable $\mathbf{v}_r$ with mean $\mathbf{As}_r$ and variance ${\sigma}^2\mathbf{I}$.
By splitting $\mathbf{s}_r$ into two identical variables $\mathbf{s}_{1,r}=\mathbf{s}_{2,r}$, an equivalent factorization can be obtained as follows
\begin{equation}\label{x12}
  p(\mathbf{v}_r,\mathbf{s}_{1,r},\mathbf{s}_{2,r})=p(\mathbf{s}_{1,r})\delta (\mathbf{s}_{1,r}-\mathbf{s}_{2,r})\mathcal{CN}(\mathbf{v}_r;\mathbf{As}_{2,r},{\sigma}^2\mathbf{I}).
\end{equation}
As specified in Algorithm 1, through passing messages on this factor according to the rules introduced in \cite{vampr}, the VAMP can be divided into a denoising step (steps $3-6$ in Algorithm 1) and a LMMSE estimation step (steps $7-10$ in Algorithm 1), which operate repeatedly for iterations $t=0,1,2,...$ until the algorithm converges.

In Algorithm 1, $\hat{\mathbf{s}}_{1,r}^t$ and $\hat{\mathbf{s}}_{2,r}^t$ are the estimates of ${\mathbf{s}}_{1,r}^t$ and ${\mathbf{s}}_{2,r}^t$ at the $t$th iteration, $\gamma _{1,r}^t$ and $\gamma _{2,r}^t$ are noise precisions, $\mathbf{u}_{1,r}^t$ and $\mathbf{u}_{2,r}^t$ behave like an AWGN corrupted version of the true signal ${\mathbf{s}}_{r}$, namely $\mathbf{u}_{1,r}^t={\mathbf{s}}_{r}+\mathcal{CN}(\mathbf{0},\mathbf{I}/\gamma _{1,r}^t)$ and $\mathbf{u}_{2,r}^t={\mathbf{s}}_{r}+\mathcal{CN}(\mathbf{0},\mathbf{I}/\gamma _{2,r}^t)$. $\mathbf{g}_{1,r}(\cdot)$ is the denoiser: $\mathbb{C}^{N\times 1}\rightarrow \mathbb{C}^{N\times 1}$, and $\mathbf{g}_{2,r}(\cdot)$ is the LMMSE estimator of $\mathbf{s}_{2,r}$. $\mathbf{g}_{1,r}^{'}(\cdot)$ and $\mathbf{g}_{2,r}^{'}(\cdot)$ are the first-order derivative of $\mathbf{g}_{1,r}(\cdot)$ and $\mathbf{g}_{2,r}(\cdot)$, respectively, and $<\cdot> $ denotes the empirical averaging operation.

\begin{algorithm}[h]
\caption{Vector AMP Algorithm}
\label{alg1}
\begin{algorithmic}[1]
\STATE \textbf{Initialize}: $\mathbf{u}_{1,r}^{0}$ and $\gamma _{2,r}^0$
\FOR{$t=0,\cdots,T_{\max}$}
\STATE $\forall r: \hat{\mathbf{s}}_{1,r}^t=\mathbf{g}_{1,r}(\mathbf{u}_{1,r}^t,\gamma _{1,r}^t)$
\STATE $\forall r: 1/\eta _{1,r}^t=\left \langle \mathbf{g}_{1,r}^{'}(\mathbf{u}_{1,r}^t,\gamma _{1,r}^t) \right \rangle/\gamma _{1,r}^t$
\STATE $\forall r: \gamma _{2,r}^t=\eta _{1,r}^t-\gamma _{1,r}^t$
\STATE $\forall r: \mathbf{u}_{2,r}^t=(\eta _{1,r}^t\hat{\mathbf{s}}_{1,r}^t-\gamma _1^t\mathbf{u}_{1,r}^t)/\gamma _{2,r}^t$
\STATE  $\forall r: \hat{\mathbf{s}}_{2,r}^t=\mathbf{g}_{2,r}(\mathbf{u}_{2,r}^t,\gamma _{2,r}^t)$
\STATE $\forall r: 1/\eta _{2,r}^t=\left \langle \mathbf{g}_{2,r}^{'}(\mathbf{u}_{2,r}^t,\gamma _{2,r}^t) \right \rangle/\gamma _{2,r}^t$
\STATE $\forall r: \gamma _{1,r}^{t+1}=\eta _{2,r}^t-\gamma _{2,r}^t$
\STATE $\forall r: \mathbf{u}_{1,r}^{t+1}=(\eta _{2,r}^t\hat{\mathbf{s}}_{2,r}^t-\gamma _{2,r}^t\mathbf{u}_{2,r}^t)/\gamma _{1,r}^{t+1}$
\ENDFOR
\end{algorithmic}
\end{algorithm}

It is shown in \cite{vampr} that when $\mathbf{A}$ is a right-orthogonally invariant random matrix and sufficiently large, the per-iteration behavior of the denoiser and the LMMSE estimator in VAMP can be exactly predicted by a scalar state evolution. The right-orthogonally invariant matrix $\mathbf{A}$ allows arbitrary singular values and arbitrary left singular vectors, making VAMP robust against the distribution of $\mathbf{A}$. However, there are still two problems when it is used for massive device detection. First, the conventional VAMP algorithm assumes that the distribution parameters are known and assumes that the device state matrix is a Bernoulli-Gaussian distributed random matrix. However, in practical massive MIMO systems \cite{channel}-\cite{channel3}, the device state matrix is a Bernoulli-Gaussian mixture distributed random matrix. As a result, the denoiser of the VAMP algorithm is not specifically designed for the massive device detection under investigation. Second, AWGN precision in $\mathbf{u}_{1,r}$ is not equal to $\boldsymbol{\gamma} _{1}$ when the distribution parameters are not perfect, which may lead to severe performance loss in VAMP algorithm.

\subsection{Bernoulli-Gaussian Mixture Distribution}
For solving the aforementioned problem, we consider a practical case that prior distribution parameters of the device state matrix $\mathbf{X}$ are unknown, which is modeled as a Bernoulli-Gaussian mixture distributed random matrix. Note that this is a general and accurate distribution for the practical scenarios with a large-scale antenna array at the BS \cite{channel}. Since the transformed device state matrix $\mathbf{S}$ has the same distribution parameters with $\mathbf{X}$, the distribution of the transformed device state matrix can be specifically expressed as
\begin{equation}\label{x}
p(\mathbf{S})=\prod_{n=1}^{N}\prod_{r=1}^{r^e}(1-\epsilon_{nr}  )\delta(s_{nr})+\epsilon_{nr} \sum\limits_{j = 1}^J {q_{nj}\mathcal{CN}({s_{nr}};0,\vartheta_{nj}^2)},
\end{equation}
where $\delta(\cdot)$ is the Dirac delta function, $0<\epsilon_{nr}<1$ is the sparse ratio, i.e., the probability of $s_{nr}$ being non-zero, $\vartheta_{nj}^2$ denotes the variance of the $j$th component, $q_{nj}>0$ is the $j$th weighted coefficient, and $\sum_{j=1}^{J}q_{nj}=1$ with $J$ being the number of the mixture components. Thus, the transformed device state matrix $\mathbf{S}$ is parameterized by unknown parameters $\epsilon_{nr}$ and $\boldsymbol{\Omega}=[\boldsymbol{\theta}_1^T,\boldsymbol{\theta}_2^T,\cdots,\boldsymbol{\theta}_n^T, \cdots, \boldsymbol{\theta}_N^T]^{T}$ with $\boldsymbol{\theta}_n=[q_{n1}, q_{n2},\cdots,q_{nJ}, \vartheta_{n1}^2,\vartheta_{n2}^2,\cdots,\vartheta_{nJ}^2]$. Note that the distribution of element in the device state matrix, $x_{nm}$, and the element in the transformed device state matrix, $s_{nr}$, have the same number of the mixture components $J$. The unknown parameters in the form of $\boldsymbol{\Omega}$ control the accuracy and convergence of the activity detection algorithm. Although dimension reduction is performed, finding optimal distribution parameters still has high computational complexity. In fact, these parameters can be learned via the back-propagation with a finite number of training data, which will be introduced in Section IV. E.

With such a Bernoulli-Gaussian mixture distribution model, we propose a deep learning-based minimum mean square error (MMSE) estimator to solve the problem of the transformed device state matrix recovery. With the low-dimensional received signal $\mathbf{V}$ and the learned variables $\hat{\boldsymbol{\Omega}}$, the MMSE estimate of $\mathbf{S}$ can be computed as
\begin{equation}\label{xesti}
 \hat{\mathbf{S}}=\mathbb {E}[\mathbf{S}|\mathbf{V};\hat{\boldsymbol{\epsilon}},\hat{\boldsymbol{\Omega} }],
\end{equation}
where $\hat{\boldsymbol{\epsilon}}$ is the estimation of $\boldsymbol{\epsilon}$, which is the collection of elements $\epsilon_{nr}$ across $r=1,2,\cdots,r^e$ and $n=1,2,\cdots,N$, and the expectation is taken over the following posterior density
\begin{equation}\label{pyc}
  p(\mathbf{S}|\mathbf{V};\hat{\boldsymbol{\epsilon}},\hat{\boldsymbol{\Omega} })=\frac{p(\mathbf{S};\hat{\boldsymbol{\epsilon}},\hat{\boldsymbol{\Omega} })p(\mathbf{V|S};\hat{\boldsymbol{\epsilon}},\hat{\boldsymbol{\Omega} })}{p(\mathbf{V};\hat{\boldsymbol{\epsilon}},\hat{\boldsymbol{\Omega} })},
\end{equation}
where $p(\mathbf{V}|\mathbf{S};\hat{\boldsymbol{\epsilon}},\hat{\boldsymbol{\Omega} })$ and $p(\mathbf{S};\hat{\boldsymbol{\epsilon}},\hat{\boldsymbol{\Omega} })$ in Eq. (\ref{pyc}) can be decoupled across the columns of ${\mathbf{S}}$ and $\mathbf{V}$ as $p(\mathbf{V}|\mathbf{S};\hat{\boldsymbol{\epsilon}},\hat{\boldsymbol{\Omega} })=\prod_{r=1}^{r^e}p(\mathbf{v}_r|\mathbf{s}_r;\hat{\boldsymbol{\epsilon}},\hat{\boldsymbol{\Omega} })$ and $p(\mathbf{S};\hat{\boldsymbol{\epsilon}},\hat{\boldsymbol{\Omega} })=\prod_{r=1}^{r^e}p(\mathbf{s}_r;\hat{\boldsymbol{\epsilon}},\hat{\boldsymbol{\Omega} })$.

Clearly, we are interested in computing the MMSE estimate of $\mathbf{S}$ from the noisy data $\mathbf{V}$. In this context, we need to design a specific denoiser for Algorithm 1 based on \eqref{xesti} and provides appropriate parameters $\hat{\boldsymbol{\Omega}}$ and $\hat{\boldsymbol{\epsilon}}$ for updating the estimated transformed device state matrix.

\subsection{The Specific Denoiser Design}
Now, the denoiser in step $3$ and its derivative in step $4$ of Algorithm 1 can be specifically designed based on \eqref{x}-\eqref{pyc} for massive device detection.
According to the theory of statistical signal processing, the element-wise Bernoulli-Gaussian mixture denoiser for $\hat{{s}}_{1,nr}$ in Algorithm 1 based on the MMSE principle is calculated as
\begin{align}\label{e1}
&\hat{{s}}_{1,nr}={g}_{1,nr}(u_{1,nr};\epsilon_{nr},\gamma _{1,r},\boldsymbol{\theta}_n)=\mathbb{E}[s_{nr}|u_{1,nr};\epsilon_{nr},\gamma _{1,r},\boldsymbol{\theta}_n]\nonumber\\
&=\frac{\int s_{nr}p\left(u_{1,nr}|s_{nr};\frac{1}{\gamma _{1,r}}\right)p(s_{nr};\epsilon_{nr},\boldsymbol{\theta}_n)ds_{nr}}{\int p\left(u_{1,nr}|s_{nr};\frac{1}{\gamma _{1,r}}\right)p(s_{nr};\epsilon_{nr},\boldsymbol{\theta}_n)ds_{nr} },
\end{align}
where ${u}_{1,nr}$ denotes the $n$th element of the vector $\mathbf{u}_{1,r}$. For the VAMP algorithm, the input of denoiser, i.e., $u_{1,nr}$, can be modeled as
\begin{align}\label{inp}
u_{1,nr}=s_{nr}+z,
\end{align}
where $z$ is the AWGN scalar which follows the distribution $\mathcal{CN}(0,1/\gamma _{1,r})$ \cite{vampr}. Consequently, we have
\begin{equation}\label{e2}
  p\left(u_{1,nr}|s_{nr};\frac{1}{\gamma _{1,r}}\right)=\mathcal{CN}\left(u_{1,nr};s_{nr},\frac{1}{\gamma _{1,r}}\right).
\end{equation}
By the law of distribution in \eqref{x}, the distribution of ${s}_{nr}$ can be written as
\begin{equation}\label{e3}
    p({s}_{nr};\epsilon_{nr},\boldsymbol{\theta}_n) =(1-\epsilon_{nr}  )\delta(s_{nr})+\epsilon_{nr}  \sum\limits_{j = 1}^J {{q_{nj}}\mathcal{CN}({s_{nr}};0,\vartheta_{nj}^2)}.
\end{equation}
Substituting the distributions defined in (\ref{e2}) and (\ref{e3}) into the expression $p\left(u_{1,nr}|s_{nr};\frac{1}{\gamma _{1,r}}\right)p(s_{nr};\epsilon_{nr},\boldsymbol{\theta}_n)$ and utilizing some calculations, we obtain
\begin{align}\label{e5}
&p\left(u_{1,nr}|s_{nr};\frac{1}{\gamma _{1,r}}\right)p(s_{nr};\epsilon_{nr},\boldsymbol{\theta}_n)\nonumber \\
&=\left((1-\epsilon_{nr}  )\delta(s_{nr})+\epsilon_{nr}  \sum\limits_{j = 1}^J {{q_{nj}}\mathcal{CN}({s_{nr}};0,\vartheta_{nj}^2)} \right )\nonumber\\ &\cdot\mathcal{CN}\left(u_{1,nr};s_{nr},\frac{1}{\gamma _{1,r}}\right) \nonumber \\
&=\epsilon_{nr}\sum\limits_{j = 1}^J {{q_{nj}}\mathcal{CN}\left({u_{1,nr}};0,\vartheta_{nj}^2\!+\!\frac{1}{\gamma _{1,r}}\right)} \nonumber\\
&\cdot\mathcal{CN}\left(s_{nr};\frac{\vartheta_{nj}^2u_{1,nr}}{b_{nj}},\frac{ \frac{1}{\gamma _{1,r}}\vartheta_{nj}^2}{b_{nj}}\right)\nonumber\\
&+(1\!-\!\epsilon_{nr}  )\delta(s_{nr})\mathcal{CN}\left(u_{1,nr};s_{nr},\frac{1}{\gamma _{1,r}}\right),
\end{align}
with
\begin{equation}\label{c2}
  b_{nj}=\vartheta_{nj}^2+\frac{1}{\gamma _{1,r}}.
\end{equation}
By substituting (\ref{e5}) into (\ref{e1}), the denoiser based on the Bernoulli-Gaussian mixture distribution can be expressed as
\begin{align}\label{e6}
{g}_{1,nr}(u_{1,nr};\epsilon_{nr},\gamma _{1,r},\boldsymbol{\theta}_n)=\frac{\epsilon_{nr}\sum\limits_{j = 1}^J {{q_{nj}}\frac{\vartheta_{nj}^2u_{1,nr}}{b_{nj}}C_{0,nj}}}{\epsilon_{nr}\sum\limits_{j = 1}^J {{q_{nj}}C_{0,nj}}+(1-\epsilon_{nr})C_{1,n}},
\end{align}
with
\begin{equation}\label{c0}
 C_{0,nj}=\mathcal{CN}({u_{1,nr}};0,b_{nj}),
\end{equation}
and
\begin{equation}\label{c1}
C_{1,n}=\mathcal{CN}\left(u_{1,nr};0,\frac{1}{\gamma _{1,r}}\right).
\end{equation}
Then, the first-order derivative of ${g}_{1,nr}(\cdot)$ can be derived as
\begin{align}\label{e7}
&{g}_{1,nr}^{'}(u_{1,nr};\epsilon_{nr},\gamma _{1,r},\boldsymbol{\theta}_n)\nonumber\\
&=\frac{\epsilon_{nr}\left(\sum\limits_{j = 1}^J {{q_{nj}}\frac{\vartheta_{nj}^2}{b_{nj}}C_{0,nj}}-\sum\limits_{j = 1}^J {{q_{nj}}\frac{\vartheta_{nj}^2u_{1,nr}}{b_{nj}}C_{0,nj}\frac{u_{1,nr}}{b_{nj}}}\right)}{\epsilon_{nr}\sum\limits_{j = 1}^J {{q_{nj}}C_{0,nj}}+(1-\epsilon_{nr})C_{1,n}}\nonumber \\
&+\frac{\sum\limits_{j = 1}^J {\frac{q_{nj}\epsilon_{nr}\vartheta_{nj}^2u_{1,nr}C_{0,nj}}{b_{nj}}}}{\left( \epsilon_{nr}\sum\limits_{j = 1}^J {{q_{nj}}C_{0,nj}}+(1-\epsilon_{nr})C_{1,n}\right)^2}\nonumber\\
&\cdot\left(\sum\limits_{j = 1}^J {\frac{q_{nj}\epsilon_{nr}u_{1,nr}C_{0,nj}}{b_{nj}}} +\frac{(1-\epsilon_{nr})u_{1,nr}C_{1,n}}{\frac{1}{\gamma _{1,r}}}\right ).
\end{align}
Compared to the existing Gaussian mixture denoiser \cite{shao1, AMP}, the proposed Bernoulli-Gaussian mixture denoiser takes into account the prior knowledge of the device state matrix, which can further improve the detection performance. In the following, we will design a FAT-DL network based on \eqref{e6} and \eqref{e7}.

\subsection{Adaptive-Tuning Module Design for Network}
To further improve the performance, we design an adaptive-tuning module by combing EM and back-propagation to adaptively tune the noise precision and learn the parameters in \eqref{xesti}. As mentioned in IV. A, $\mathbf{u}_{1,r}^t$ behaves like a corrupted version of the true signal ${\mathbf{s}}_{r}$ by an AWGN precision $\gamma _{1,r}^t$, namely $\mathbf{u}_{1,r}^t={\mathbf{s}}_{r}+\mathcal{CN}(\mathbf{0},\mathbf{I}/\gamma _{1,r}^t)$. When the variable $\boldsymbol{\Omega}$ learned by the deep learning network is not perfect, $\mathbf{u}_{1,r}^t$ is still an AWGN corrupted version of the true signal ${\mathbf{s}}_{r}$. In this case, the AWGN precision is not equal to $\gamma _{1,r}^t$, which compromises the updates of $\hat{\mathbf{s}}_{1}^t$.
Thus, we propose to adaptively tune the precision $\boldsymbol{\gamma}_{1}^t=[\gamma _{1,1}^t,\cdots,\gamma _{1,r^e}^t]^T$ by combing back-propagation. In this context, the maximum likelihood (ML) estimator is employed, which is given by
\begin{equation}\label{ga1}
  \{\boldsymbol{\gamma}_1^t,\boldsymbol{\epsilon}^t\}=\mathop \text{argmax}\limits_{\boldsymbol{\gamma}_1}p(\mathbf{U}_1^t;\boldsymbol{\epsilon}, \boldsymbol{\gamma}_1,\mathbf{\Omega}),
\end{equation}
under the statistical model of $\mathbf{u}_{1,r}^t$ and the distribution feature of ${\mathbf{s}}_{r}$. Therein, $\boldsymbol{\Omega}$ is learned through back-propagation and $\mathbf{U}_1^t$ is the matrix collecting $\mathbf{u}_{1,r}^t$. To solve the problem (\ref{ga1}), we utilize the following inner EM iterations based on the learned $\mathbf{\Omega}^{\tau}$ in the $\tau$th layer
\begin{align}
\label{em}
&\boldsymbol{\gamma}_1^{\tau+1}=\mathop \text{argmax} \limits_{\boldsymbol{\gamma}  _1}\mathbb{E}[\ln p(\mathbf{S},\mathbf{U}_1^t;\boldsymbol{\epsilon},\boldsymbol{\gamma}_1,\mathbf{\Omega})|\mathbf{U}_1^t;\boldsymbol{\epsilon}^{\tau},\boldsymbol{\gamma}_1^{\tau},\mathbf{\Omega}^{\tau}]\nonumber\\
&=\mathop \text{argmax} \limits_{\boldsymbol{\gamma}  _1}\{\mathbb{E}[\ln p(\mathbf{S};\boldsymbol{\epsilon},\mathbf{\Omega})|\mathbf{U}_1^t;\boldsymbol{\epsilon}^{\tau},\boldsymbol{\gamma}_1^{\tau},\mathbf{\Omega}^{\tau}]\nonumber\\
&+ \mathbb{E}[\ln p(\mathbf{U}_1^t|\mathbf{S};\boldsymbol{\gamma} _{1})|\mathbf{U}_1^t;\boldsymbol{\epsilon}^{\tau},\boldsymbol{\gamma}_1^{\tau},\mathbf{\Omega}^{\tau}]\}.
\end{align}
Keeping the variables of outer iteration $t$ unchanged, and omitting the terms that are independent of the parameter vector $\boldsymbol{\gamma}_{1}^{\tau}$, we have
\begin{equation}\label{bgam}
  \boldsymbol{\gamma}_{1}^{\tau+1}=\mathop \text{argmax} \limits_{\boldsymbol{\gamma} _{1}}\sum_{r=1}^{r^e} \mathbb{E}[\ln p(\mathbf{u}_{1,r}^t|\mathbf{s}_{r};\mathbf{\gamma }_{1,r})|\mathbf{u}_{1,r}^t;\boldsymbol{\epsilon}^{\tau},\gamma_{1,r}^{\tau},\mathbf{\Omega}^{\tau}]. \nonumber
\end{equation}
Decoupling $\boldsymbol{\gamma}_{1}^{\tau}$ into the elementwise form $\gamma_{1,r}^{\tau}$ leads to
\begin{align}
\label{gamm1}
&\gamma_{1,r}^{\tau+1}\nonumber\\
&=\mathop \text{argmax} \limits_{\gamma_{1,r}}\!\!\left\{\!\!\frac{N\ln\gamma_{1,r}\!-\!\gamma_{1,r}\mathbb{E}\left[\left \| \mathbf{s}_{r}\!\!-\!\!\mathbf{u}_{1,r}^t \right \|_2^2|\mathbf{u}_{1,r}^t;\boldsymbol{\epsilon}^{\tau},\gamma _{1,r}^{\tau},\mathbf{\Omega}^{\tau}\right]}{2}\!\! \right\}\nonumber \\
&\!\!=\!\!N\left( \mathbb{E}\left[\left \| \mathbf{s}_{r}-\mathbf{u}_{1,r}^t \right \|_2^2|\mathbf{u}_{1,r}^t;\boldsymbol{\epsilon}^{\tau},\gamma _{1,r}^{\tau},\mathbf{\Omega}^{\tau}\right]  \right)^{-1}\nonumber\\
&=\left(\frac{1}{N}\left \|\hat{\mathbf{s}}_{1,r}^{\tau}-\mathbf{u}_{1,r}^t\right \|_2^2+1/\eta _{1,r}^{\tau}\right)^{-1}
\end{align}
where $\hat{\mathbf{s}}_{1,r}^{\tau}$ and $1/\eta _{1,r}^{\tau}$ can be calculated by (\ref{d}) and (\ref{d1}), respectively. In a similar way, $\boldsymbol{\epsilon}$ can be obtained by
\begin{align}\label{epis}
\boldsymbol{\epsilon}^{\tau+1}=\mathop \text{argmax} \limits_{\boldsymbol{\epsilon}}\mathbb{E}[\ln p(\mathbf{S};\boldsymbol{\epsilon},\mathbf{\Omega})|\mathbf{U}_1^t;\boldsymbol{\epsilon}^{\tau},\boldsymbol{\gamma}_1^{\tau},\mathbf{\Omega}^{\tau}]\}.
\end{align}

For the $\hat{\mathbf{s}}_{2,r}$ in the LMMSE estimator, message passing rules have shown that it is the expectation of
$\hat{\mathbf{s}}_{2,r} \sim \mathcal{CN}(\mathbf{s}_2;\mathbf{u}_{2,r},\gamma_{2,r}^{-1}\mathbf{I} ) \mathcal{CN}(\mathbf{v}_r;\mathbf{A}\mathbf{s}_2,{\sigma}^2\mathbf{I})$. Herein, the pilot matrix $\mathbf{A}$ is updated as $\mathbf{A}^t\triangleq\mathbf{A}\text{diag}(\boldsymbol{\beta}^t)$ at the $t$th layer for a more accurate estimation by introducing the parameter $\boldsymbol{\beta}^t=[\beta_1^t,\beta_2^t,\cdots,\beta_N^t]$. In other words, the parameter $\beta_n^t$ acts to scale the power of the $n$th column of pilot matrix $\mathbf{A}$.
Compared with the scaling rule of pilot matrix in the learned AMP (LAMP) \cite{lamp}, where all elements of matrix $\mathbf{A}$ vary with the layer $t$, the proposed scheme can decrease the expense of an $L$-fold in memory and training complexity.

Using standard expectation manipulations, we have
\begin{align}\label{s2}
  \hat{\mathbf{s}}_{2,r}&=\left(\frac{1}{{\sigma}^2}\text{diag}(\boldsymbol{\beta})^H\mathbf{A}^H\mathbf{A}\text{diag}(\boldsymbol{\beta})+
  \gamma _{2,r}
  \mathbf{I}_N\right)^{-1}\nonumber\\
  &\cdot\left(\gamma _{2,r}\mathbf{u} _{2,r}+\frac{1}{\sigma^2}\text{diag}(\boldsymbol{\beta})^H\mathbf{A}^H\mathbf{v}_r\right).
\end{align}
Taking the derivative over the equation above and performing the empirical averaging leads to
\begin{align}\label{oop}
&\left \langle \mathbf{g}_{2,r}^{'}(\mathbf{u}_{2,r};\gamma _{2,r},\boldsymbol{\beta}) \right \rangle \nonumber\\
&=\frac{\gamma_{2,r}}{N}\text{tr}\left[\left(\frac{1}{\sigma^2}\text{diag}(\boldsymbol{\beta})^H\mathbf{A}^H\mathbf{A}\text{diag}(\boldsymbol{\beta})+\gamma_{2,r}\mathbf{I}\right)^{-1}\right].
\end{align}

In the following, we will propose a deep learning-based JADCE network based on the designed denoiser and adaptive-tuning module for jointly learning the unknown distribution parameters and applying the learned distribution parameters to detect active devices. Specifically, the designed FAT-DL network contains an inner network and an outer network. The outer network consists of $T_{\max}$ cascade layers and each has the same architecture but different trainable parameters. For the $t$-th layer of the outer network, the input is the estimation $\mathbf{u}_{1,r}^t$ and $\gamma_{1,r}^{{t}}$ from the $(t-1)$th layer, and the output variables are $\mathbf{u}_{1,r}^{t+1}$ and $\gamma_{1,r}^{t+1}$.  The inner network consists of $\tau_{\max}$ cascade layers and each has the same architecture but different trainable parameters. For the $\tau$-th layer of the inner network, the input is the estimation $\gamma_{1,r}^{{\tau}}$ from the $(\tau-1)$th layer, and the output variables is $\gamma_{1,r}^{{\tau+1}}$. In summary, the $t$th outer layer of the FAT-DL network performs as \eqref{d}-\eqref{last}, where the inner layer repeats (\ref{d})-(\ref{d3}) for $\tau = 1,\dots,\tau_{\max}$
\begin{align}
&\hat{s}_{1,nr}^{\tau}= \frac{\epsilon_{nr}^{\tau}\sum\limits_{j = 1}^J {{q_{nj}^{\tau}}\frac{\vartheta_{nj}^{2,t}{u}_{1,nr}^t}{b_{nj}^{\tau}}C_{0,nj}^{\tau}}}{\epsilon_{nr}^{\tau}\sum\limits_{j = 1}^J {{q_{nj}^{\tau}}C_{0,nj}^{\tau}}+(1-\epsilon_{nr}^{\tau})C_{1,n}^{\tau}}, \forall r, n,\label{d}\\
& 1/\eta _{1,r}^{\tau}= \langle {g}_{1,nr}^{'}(\mathbf{u}_{1,r}^t;\epsilon_{nr}^{\tau},\gamma _{1,r}^{\tau},\boldsymbol{\theta}_n^{\tau})\rangle/\gamma _{1,r}^{\tau}, \forall r, n,\label{d1}\\
& \gamma_{1,r}^{{\tau}}= \left(\frac{1}{N}\left \| \hat{\mathbf{s}}_{1,r}^{{\tau}}-\mathbf{u}_{1,r}^t \right \|_2^2+1/\eta _{1,{r}}^{\tau}\right)^{-1}, \forall r,\label{d2}\\
&\epsilon_{nr}^{\tau}=\frac{1}{r^e}\sum_{r=1}^{r^e}\frac{1}{1+\frac{(1-\epsilon_{nr}^{\tau})C_{1,n}^{\tau}}{\epsilon_{nr}^{\tau}\sum\limits_{j = 1}^J {{q_{nj}^t}C_{0,nj}^{\tau}}}}, \forall r, n,\label{ratio}\\
& \tau = \tau+1,\label{d3}
\end{align}
and then let
\begin{align}
&\hat{\mathbf{S}}_{1}^t=\hat{\mathbf{S}}_{1}^{\tau},\\
& \gamma_{1,r}^{t}=\gamma_{1,r}^{\tau}, \forall r,\\
& \gamma _{2,r}^{t}=\eta _{1,r}^t-\gamma _{1,r}^t, \forall r,\\
& \mathbf{u}_{2,r}^{t}=(\eta _{1,r}^t\hat{\mathbf{s}}_{1,r}^t-\gamma _1^t\mathbf{u}_{1,r}^t)/\gamma _{2,r}^t, \forall r.
\end{align}
For the LMMSE estimation step:
\begin{align}
& \hat{\mathbf{s}}_{2,r}^t=\left(\frac{1}{{\sigma}^2}\text{diag}(\boldsymbol{\beta}^t)^H\mathbf{A}^H\mathbf{A}\text{diag}(\boldsymbol{\beta}^t)\!+\!
  \gamma _{2,r}^t
  \mathbf{I}_N\right)^{-1}\nonumber\\
  &~~~~~\cdot\left(\gamma _{2,r}^t\mathbf{u} _{2,r}^t\!+\!\frac{1}{\sigma^2}\text{diag}(\boldsymbol{\beta}^t)^H\mathbf{A}^H\mathbf{v}_r\right), \forall r, \label{s2r}\\
& \eta _{2,r}^{t}= \frac{1}{N}\text{tr}\left[\left(\frac{1}{\sigma^2}\mathbf{A}(\boldsymbol{\beta}^t)^H\mathbf{A}(\boldsymbol{\beta}^t)+\gamma_{2,r}^{t}\mathbf{I}\right)^{-1}\right], \forall r, \label{i2r}\\
& \gamma _{1,r}^{t+1}=\eta _{2,r}^t-\gamma _{2,r}^t, \forall r,\\
& \mathbf{u}_{1,r}^{t+1}=(\eta _{2,r}^t\hat{\mathbf{s}}_{2,r}^t-\gamma _{2,r}^t\mathbf{u}_{2,r}^t)/\gamma _{1,r}^{t+1}, \forall r.\label{last}
\end{align}

\begin{figure*}
\centering
\includegraphics [width=1.00\textwidth] {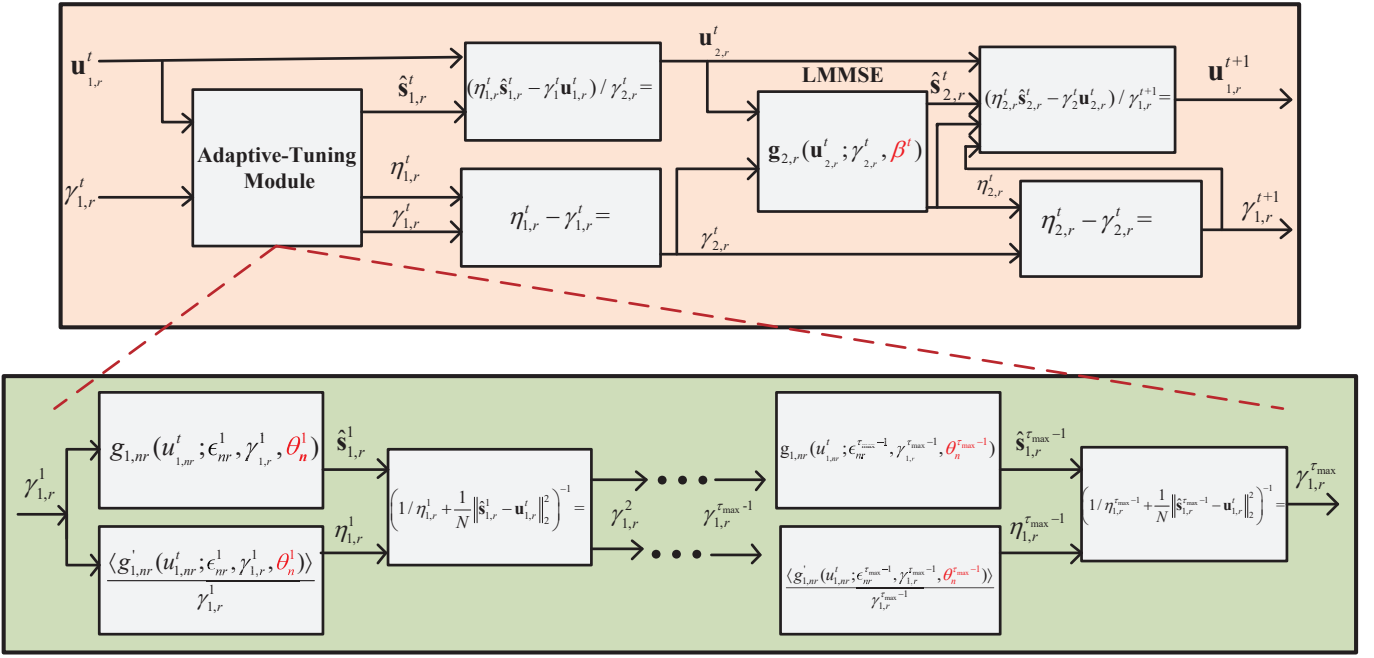}
\caption{The proposed FAT-DL architecture with the learnable variables $\boldsymbol{\beta}^t$ and $\boldsymbol{\Omega}^{\tau}=[(\boldsymbol{\theta}_1^{\tau})^T,(\boldsymbol{\theta}_2^{\tau})^T,\cdots,(\boldsymbol{\theta}_N^{\tau})^T]^{T}$. Wherein, the adaptive-tuning module and LMMSE estimator exchange information iteratively until convergence.}
\label{RV}
\end{figure*}

Notice that the rows of the transformed device state matrix $\mathbf{S}$ share the common support, which can be utilized to improve the accuracy of the support estimation.
Therefore, the update rule of the sparsity ratio in \eqref{ratio} is refined to leverage the structured sparsity of the transformed device state matrix for enhancing detection performance. Fig. \ref{RV} provides an illustration of the corresponding block diagram of the proposed FAT-DL network. The proposed network architecture is obtained by introducing the learnable variables $\mathbf{\Phi}=[\boldsymbol{\beta}, \boldsymbol{\Omega}]$ and adaptive-tuning module, i.e., the inner network, which fully exploit the domain knowledge and adaptively adjust the AWGN precision. This improved architecture not only inherits the superiority of traditional approaches, but also significantly
improves the performance of activity detection.

\subsection{Parameter Learning}
We now turn to design a scheme to train the parameters $\boldsymbol{\Omega}$ characterizing the Bernoulli-Gaussian mixture distribution of the device state matrix, and the parameters $\boldsymbol{\beta}$ adjusting the pilot matrix for the designed deep learning network. From Fig. \ref{RV}, we observe that the convergence behavior and detection performance of the designed algorithm are determined by appropriate parameters $\boldsymbol{\Omega}$. The parameters $\boldsymbol{\beta}$ in the LMMSE estimation step also play important roles in constructing transform coefficients.

Before proceeding, we introduce the inner loss function $C_{\text{Inner}}^{t}$ and outer loss function $C_{\text{Outer}}^{t}$ as follows:
\begin{equation}\label{inner}
  C_{\text{Inner}}^{t}=\left \| \hat{\mathbf{S}}_{1}^t-\mathbf{S} \right \|_F^2,
\end{equation}
\begin{equation}\label{outer}
  C_{\text{Outer}}^{t}=\left \| \hat{\mathbf{S}}_{2}^t-\mathbf{S} \right \|_F^2,
\end{equation}
where $\mathbf{S}$ is the true transformed device state matrix, $\hat{\mathbf{S}}_{1}^t$ and $\hat{\mathbf{S}}_{2}^t$ are the matrices collecting $\hat{\mathbf{s}}_{1,r}^t$ and $\hat{\mathbf{s}}_{2,r}^t$, respectively.

The details of the parameters learning are specified in Algorithm 2. The training includes an inner training by minimizing the inner loss function $C_{\text{Inner}}^{t}$ and a outer training by minimizing the outer loss function $C_{\text{Outer}}^{t}$, where coefficients $\boldsymbol{\beta}^t$ or distribution parameters $\boldsymbol{\Omega}^i$ are first optimized individually, and then
all the trainable variables of the previous $t$ outer layers and $i$ inner layers are optimized globally. Specifically, line $14$ performs a training by minimizing the inner loss function $C_{\text{Inner}}^{i}$, where only the $\boldsymbol{\Omega}^i$ is trainable with the trainable variables $\boldsymbol{\beta}^{t}$ of previous $(t-1)$ outer layers and $\boldsymbol{\Omega}^i$ of previous $(i-1)$ inner layers unchanged. Next, line $15$ re-learns the inner parameters $\{\boldsymbol{\Omega}^i\}_{i=t\tau_{\max}-\tau_{\max}+1}^{i}$ for the training limited in the layer $t$.
Line $17$ re-learns the variables $\boldsymbol{\Omega}^i$ of previous $i$ inner layers and variables $\boldsymbol{\beta}^t$ of previous $(t-1)$ outer layers. Line $18$ performs a training by minimizing the outer loss function $C_{\text{Outer}}^{t}$ in the LMMSE estimation step, where only the $\boldsymbol{\beta}^t$ is trainable with the trainable variables $\boldsymbol{\beta}^{t}$ of previous $(t-1)$ outer layers and $\boldsymbol{\Omega}^i$ of previous $i$ inner layers unchanged. Finally, line $19$ performs the global optimization for the training.

Compared with the LVAMP network in \cite{lamp}, where the soft-threshold denoiser is applied with the trainable threshold parameter, the proposed Bernoulli-Gaussian mixture denoiser takes into account the prior knowledge of the practical transformed device state matrix, which can improve the detection performance. On the other hand, compared with the LVAMP network, an inner network is introduced to the proposed FAT-DL network, which can further reduce the performance loss caused by imperfect learning of the prior distribution parameters. Furthermore, different from the conventional deep learning network, which only defines one loss function for the whole network, the proposed deep learning network employs two loss functions in each layer, which are related to the inner network and the outer network, respectively. Finally, the calculations of the conventional LVAMP network are made on the real and imaginary part of the complex inputs separately, while the proposed FAT-DL network directly conducts the denoising step and LMMSE step on the complex inputs.

\begin{algorithm}[phtb]
\caption{FAT-DL Parameter Learning.}
\label{alg1}
\begin{algorithmic}[1]
\STATE \textbf{Initialization}: $\boldsymbol{\beta}^1=\boldsymbol{\beta}^0$, $\boldsymbol{\Omega}^1=\boldsymbol{\Omega}^0$
\STATE Learn $\boldsymbol{\Omega}^1$ to minimize $C_{\text{Inner}}^{1}$
\FOR{$i=2,\cdots,\tau_{\max}$}
\STATE Learn $\boldsymbol{\Omega}^i$ with fixed $\{\boldsymbol{\Omega}^i\}_{i=1}^{i-1}$ to minimize $C_{\text{Inner}}^{i}$
\STATE Re-learn $\{\boldsymbol{\Omega}^i\}_{i=1}^{i}$ to minimize $C_{\text{Inner}}^{i}$
\ENDFOR
\STATE Learn $\boldsymbol{\beta}^{1}$ with fixed $\{\boldsymbol{\Omega}^i\}_{i=1}^{i}$ to minimize $C_{\text{Outer}}^{1}$
\STATE Re-learn $\{\boldsymbol{\beta}^{1},\{\boldsymbol{\Omega}^i\}_{i=1}^{i}\}$ to minimize $C_{\text{Outer}}^{1}$
\FOR{$t=2,\cdots,T_{\max}$}
\STATE Initialization $\boldsymbol{\beta}^{t}=\boldsymbol{\beta}^{t-1}$
\FOR{$\tau=1,\cdots,\tau_{\max}$}
\STATE $i\leftarrow i+1$
\STATE Initialization $\boldsymbol{\Omega}^i=\boldsymbol{\Omega}^{i-1}$
\STATE Learn $\boldsymbol{\Omega}^i$ with fixed $\{\boldsymbol{\Omega}^i\}_{i=1}^{i-1}$ and $\{\boldsymbol{\beta}^{t}\}_{t=1}^{t-1}$ to minimize $C_{\text{Inner}}^{i}$
\STATE Re-learn $\{\boldsymbol{\Omega}^i\}_{i=t\tau_{\max}-\tau_{\max}+1}^{i}$ with fixed $\{\boldsymbol{\beta}^{t}\}_{t=1}^{t-1}$ and $\{\boldsymbol{\Omega}^i\}_{i=1}^{i=t\tau_{\max}-\tau_{\max}}$ to minimize $C_{\text{Inner}}^{i}$
\ENDFOR
\STATE Re-learn $\{\boldsymbol{\Omega}^i\}_{i=1}^{i}$ and $\{\boldsymbol{\beta}^{t}\}_{t=1}^{t-1}$ to minimize $C_{\text{Outer}}^{t}$
\STATE
Learn $\boldsymbol{\beta}^t$ with fixed $\{\boldsymbol{\Omega}^i\}_{i=1}^{i}$ and $\{\boldsymbol{\beta}^{t}\}_{t=1}^{t-1}$  to minimize $C_{\text{Outer}}^{t}$
\STATE
Re-learn $\{\boldsymbol{\beta}^{t}\}_{t=1}^{t}$ and $\{\boldsymbol{\Omega}^i\}_{i=1}^{i}$ to minimize $C_{\text{Outer}}^{t}$
\STATE Update $t\leftarrow t+1$
\ENDFOR
\STATE \textbf{Return}:  $\{\boldsymbol{\beta}^{t}\}_{t=1}^{T_{\max}}$ and $\{\boldsymbol{\Omega}^i\}_{i=1}^{T_{\max}\tau_{\max}}$
\end{algorithmic}
\end{algorithm}

\subsection{Implementation Details and Insights}
The proposed FAT-DL algorithm works in two phases: an offline training phase and an online JADCE phase. In the offline training phase, we obtain the parameters
$\{\boldsymbol{\beta}^{t}\}_{t=1}^{T_{\max}}$ and $\{\boldsymbol{\Omega}^i\}_{i=1}^{T_{\max}\tau_{\max}}$ for the deep learning network by minimizing the loss function with a finite number of known training data generated from \eqref{eqinter5}. In the online JADCE phase, the FAT-DL network can be regarded as an iterative estimator.  According to \eqref{eqinter5}, the new measurements different from the training data set can be obtained. Then,
the new measurements and the stored parameters $\{\boldsymbol{\beta}^{t}\}_{t=1}^{T_{\max}}$ and $\{\boldsymbol{\Omega}^i\}_{i=1}^{T_{\max}\tau_{\max}}$ obtained from the offline phase are fed into the FAT-DL network in turn to directly generate the estimation of transformed device state matrix.

\emph{Remark 1}: The inner network can be extended to the LMMSE estimation step in \eqref{s2r} and \eqref{i2r} for improving performance. For example, besides the distribution parameters, the accurate noise variance $\sigma^2$ is not trivial to get in practical applications, which can be learned by deep learning networks. Also, AWGN precision in $\mathbf{u}_{2,r}$ is not equal to $\gamma _{2,r}$ when the learned variance is not perfect, which may lead to severe performance loss in the FAT-DL algorithm.
The inner network about precision $\boldsymbol{\gamma}_{2}=[\gamma _{2,1},\cdots,\gamma _{2,r^e}]^T$ can be introduced. Accordingly, a specialized parameter learning scheme for this specific problem can be designed for better performance.

Unlike conventional deep neural network (DNN) based activity detection \cite{wei}, where the adopted activation function is a black box that has no explicit physical meaning, the designed denoiser in \eqref{e6} of the proposed FAT-DL network is equivalent to the activation function. The designed denoiser has an explicit physical meaning, this is because the denoiser can promote $\hat{\mathbf{S}}$ sparser by minimizing the mean-squared-error (MSE) in each iteration. In this context, the proposed FAT-DL network requires less low-dimensional training data and shorter training time. However, in \cite{wei}, to train the black box, it requires large training data and training time, which are often scarce in $6$G wireless networks due to the fact that the fading channel is time-varying and the activity status of the devices changes over the time slots. In addition, the proposed FAT-DL network is obtained by unfolding an
iterative algorithm, thus its performance can be rigorously analyzed through combing the specific denoiser in \eqref{e6} with the state evolution \cite{AMP}, which will be discussed in the future work.

In summary, the proposed FAT-DL algorithm can obtain a more accurate complex-valued estimation with low computational complexity and low demand for training data by efficiently learning the distribution parameters of the transformed device state matrix and boosting the adaptive update of the AWGN precision. Moreover, the proposed FAT-DL algorithm is robust to a much broader class of pilot matrices, as will be verified by simulations in Section V, which stabilizes the system and saves the storage space of the BS in the context of mMTC compared to the detectors that can only work in the case of a large i.i.d. Gaussian pilot matrix. Therefore, the proposed FAT-DL algorithm becomes appealing for achieving intelligent device detection.

\subsection{Computational Complexity}
In what follows, the computational complexity of the proposed algorithm is briefly discussed. Unlike conventional massive device detection algorithms, the proposed algorithm does not recover the original unknown signal in the high-dimensional space $\mathbb{C}^{N\times M}$. It recovers the sparse signal in a potentially low-dimensional space $\mathbb{C}^{N\times r^e}$. Specifically, the computational complexity of the FAT-DL algorithm mainly comes from the matrix multiplication $\mathbf{AS}$. Given $\mathbf{A}\in \mathbb{C}^{L\times N}$ and $\mathbf{S}\in \mathbb{C}^{N\times r^e}$, the computational complexity of FAT-DL is in the order of $\mathcal{O}(LNr^e)$ per layer. In contrast with the AMP-based device detection algorithm \cite{AMPO,AMPliang1}, the computational complexity of FAT-DL does not grow by increasing the number of BS antennas $M$.
FAT-DL learns $\boldsymbol{\Omega} \in \mathbb{C}^{N\times 2J}$ and $\boldsymbol{\beta} \in \mathbb{C}^{N\times 1}$ for inner and outer layers. Thus, its memory complexity can be approximated as $T_{\max}(N+2NJ\tau_{\max})$ over $T_{\max}$ layers.

In this paper, we compare the proposed algorithm with five traditional algorithms from the computational complexity and memory complexity aspects, including the AMP algorithm \cite{AMP}, the learned AMP (LAMP) algorithm \cite{lamp}, the learned vector AMP (LVAMP) algorithm \cite{lamp}, the fast iterative shrinkage-thresholding algorithm (FISTA) \cite{fista} which is a classical optimization algorithm to minimize convex functions, and the OMP algorithm which is a greedy algorithm proposed in \cite{OMP}. It can be seen in Table \ref{tab1} that the computational complexity scalings of FAT-DL is superior to the traditional algorithms, implying its lower complexity in the high dimensional regime. Note that, an $r^e \times r^e$ SVD is performed for dimension reduction, whose computational complexity in the worst case is $\mathcal{O}(\min(L,M)^2)$. For sparse recovery, since $r^e \leq K$ and SVD only
needs to be calculated once at each time slot before the start of iteration, this step is not numerically expensive. Moreover, the memory complexity of the proposed FAT-DL algorithm is lower than that of the LAMP and LVAMP algorithms in mMTC, which is an advantageous feature for massive device detection problems.

\newcommand{\tabincell}[2]{\begin{tabular}
{@{}#1@{}}#2\end{tabular}}
\begin{table}[h]
\centering
\caption{The Computation Complexity and Memory Complexity Comparison of Considered Algorithms.}
\label{tab1}
\begin{tabular}{lcccc}
\hline
\tabincell{c}{Algorithms}     &\tabincell{c}{Computational\\ complexity} &\tabincell{c}{Memory\\ complexity}
\\
\hline
FAT-DL & $\mathcal{O}(LNr^e)$ &$T_{\max}(N+2JN\tau_{\max})$
  \\
LAMP \cite{lamp} & $\mathcal{O}(LNM)$&$T_{\max}(LN+2)$
  \\
LVAMP \cite{lamp} & $\mathcal{O}(LNM)$&$T_{\max}(LN^2+2)$
  \\
AMP \cite{AMP} &  $\mathcal{O}(LNM)$&0\\
FISTA \cite{fista} &  $\mathcal{O}(LNM)$&0  \\
OMP \cite{OMP}&  $\mathcal{O}(LNM+K^3M^3)$&0  \\
\hline
\end{tabular}
\end{table}

\section{Numerical Results}
We examine the activity detection performance and the channel estimation accuracy of the proposed algorithm through computer simulations. As a reference, we compare the proposed FAT-DL algorithm with the AMP algorithm \cite{AMP}, the LAMP algorithm \cite{lamp}, the LVAMP algorithm \cite{lamp}, the FISTA algorithm \cite{fista}, and the OMP algorithm \cite{OMP}. We use the activity error rate (AER) to measure the detection performance and normalized mean square error (NMSE) to measure the channel estimation accuracy. The AER is a sum of the miss detection probability, defined as the probability that a device is active but is declared to be inactive, and the false-alarm probability, defined as the probability that a device is inactive but the detector declares it to be active. The NMSE of all active devices is defined as $10\log_{10}\frac{\left \| \hat{\mathbf{X}}_{\mathcal{K}}-\mathbf{X}_{\mathcal{K}} \right \|_F^2}{\left \| \mathbf{X}_{\mathcal{K}} \right \|_F^2}$ where $\mathbf{X}_{\mathcal{K}}$ collects the row vectors corresponding to the active support $\mathcal{K}$ in $\mathbf{X}$. The SNR is defined as $10\log_{10}(\left\|\mathbf{AX}\right\|_F^2/{LM\sigma^2})$.

The spatial channels of all IoT devices are generated according to \cite{spachan} and each device has different distribution parameters. For each device, the number of paths is set as $3$, the complex gain is generated from a complex-valued standard normal distribution, the azimuth and elevation of each path are generated from $(-\pi/2,\pi/2)$, $v_1$ is set to $0.1$, and the number of components $J$ is set to $3$, unless stated otherwise. For the proposed FAT-DL network, we use samples of size $100,000$ for training, $30,000$ for validation, and $30,000$ for testing. Herein, the testing data and training data follow the same distribution. The mini-batch size is set to be $128$. In particular, we randomly select $128$ samples from the training data set at each updating, and the total size of updating is set as $1,000,000$. The training and testing methods were implemented in Python using TensorFlow with the Adam optimizer \cite{tens,adam}. The training rates for individual
optimization of Algorithm 2 are set as $0.001$, and for the global optimization of Algorithm 2, the training rate decreases to $0.0005$, $0.0001$, and $0.00001$ to reduce the validation error. The configures of the LAMP and LVAMP algorithms are the same as that of the FAT-DL algorithm. For the OMP-based device detection algorithm, the number of iterations is equal to the sparsity level of the vectorized device state matrix. For the FISTA algorithm, we set the maximum number of iterations as $500$, which is enough to get a maximal absolute error inferior to an acceptable value between two iterations.

\begin{figure}[h]
\centering
\includegraphics [width=0.5\textwidth]{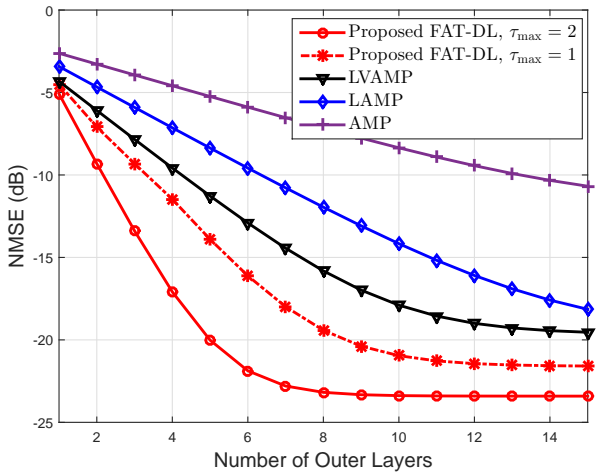}
\caption{The NMSE for different numbers of outer layers with $N = 100$, $\epsilon_{k}=0.05$, $L=60$, $M = 64$, SNR = $30$ dB, and $\mathbf{A}$ is i.i.d. Gaussian distributed with zero mean and unit variance.}
\label{iidlayer}
\end{figure}

Fig. \ref{iidlayer} illustrates the NMSE performance versus the number of outer layers under an i.i.d. Gaussian distribution $\mathbf{A}$ with zero mean and unit variance. The soft-threshold based shrinkage functions are adopted for the AMP algorithm, LAMP algorithm, and LVAMP algorithm. To clarify the impact of the proposed training steps on the detection performance compared to a simpler training process, the proposed FAT-DL algorithm is trained with varying $\tau_{\max}$ which controls the scale of the inner network. When $\tau_{\max}=1$, the proposed FAT-DL reduces to a simpler network, where the inner network degenerates. It is seen that the proposed FAT-DL algorithm with $\tau_{\max}=1$ can obtain more accurate NMSE than the AMP algorithm, LAMP algorithm, and LVAMP algorithm. Moreover, it is found that FAT-DL needs a smaller number of layers for achieving the saturated performance when $\mathbf{A}$ is i.i.d. Gaussian distributed with zero mean and unit variance. Such advantages of the proposed FAT-DL mainly benefit from that there are learnable variables $\boldsymbol{\beta}^t$ and $\boldsymbol{\Omega}^i$ in each layer, and a specifically designed denoiser. The performance of joint activity detection and channel estimation can be improved through optimizing these variables in the training process. On the other hand, when $\tau_{\max}>1$,  the proposed FAT-DL involves both inner and outer networks. Accordingly, the detection performance of FAT-DL further improves when the value of $\tau_{\max}$ becomes large. This is due to the fact that when $\tau_{\max}>1$, the AWGN precision $\boldsymbol{\gamma}_1^t$ is updated based on the current learned prior distribution parameters, hence, the inner network can further reduce the performance loss caused by imperfect learning of the prior distribution parameters.

\begin{figure}[h]
\centering
\includegraphics [width=0.5\textwidth] {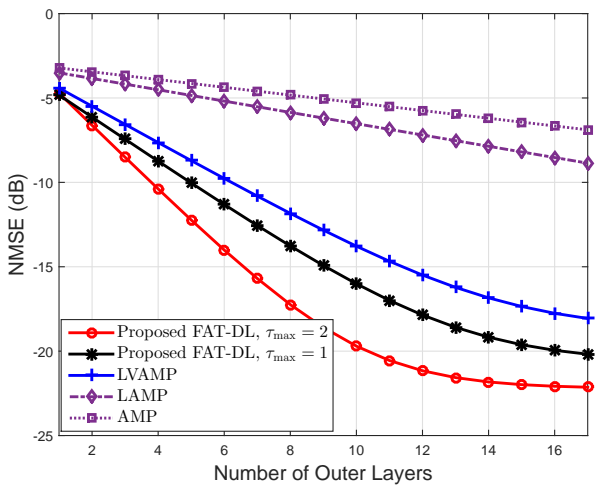}
\caption{The NMSE for different numbers of outer layers with $N = 100$, $\epsilon_{k}=0.05$, $L=60$, $M = 64$, SNR = $30$ dB, and $\mathbf{A}$ has the condition number 20.}
\label{condlayer}
\end{figure}

Fig. \ref{condlayer} shows the NMSE performance of the proposed algorithm versus the number of outer layers under pilot matrix $\mathbf{A}$ with the condition number = $20$. As seen in Fig. \ref{condlayer}, the proposed FAT-DL algorithms with $\tau_{\max}=1,2$ perform better than the AMP, LAMP, and LVAMP algorithms in the entire range of layer numbers. Note that the AMP algorithm and the LAMP algorithm are sensitive to the ill-conditioned pilot matrix. However, the FAT-DL algorithm has higher robustness than the other considered algorithms when the pilot matrix is an ill-conditioned matrix. Such advantages mainly stem from the fact that the proposed FAT-DL algorithm not only introduces the adaptive-tuning module and exploits the prior information of the Bernoulli-Gaussian mixture distributed channel via deep learning, but also inherits the advantages of the traditional VAMP algorithm effectively.

\begin{figure}[h]
\centering
\includegraphics [width=0.5\textwidth] {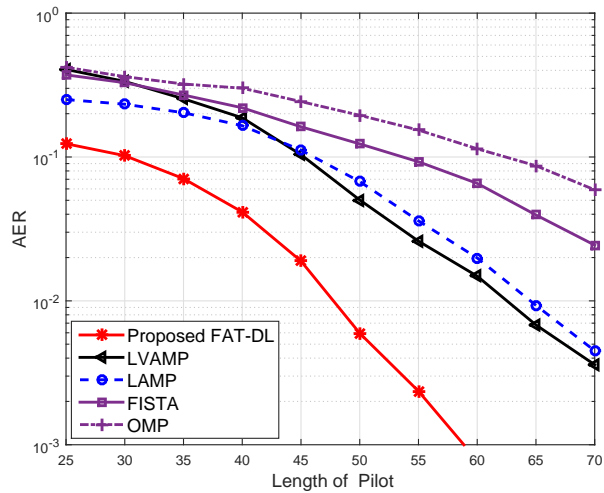}
\caption{The AER for different lengths of pilot sequence with $N = 100$, $\epsilon_{k}=0.2$, $M = 32$, SNR = $20$ dB, and $\mathbf{A}$ is i.i.d. Gaussian distributed with zero mean and unit variance.}
\label{pilotmean1}
\end{figure}

In the rest of the simulations, $\tau_{\max}$ is set to $2$ for unveiling the full potential of FAT-DL under various system settings. Fig. \ref{pilotmean1} illustrates the AER performance versus the length of pilot sequences under an i.i.d. Gaussian distribution $\mathbf{A}$ with zero mean and unit variance. It is clear that the proposed FAT-DL algorithm can achieve a better performance than the AMP algorithm, the FISTA algorithm, the OMP algorithm, the LAMP algorithm, and the LVAMP algorithm with the same length of pilot sequences. In other words, the proposed FAT-DL algorithm needs shorter pilot sequences than the conventional algorithms to achieve the same activity detection accuracy. For example, the AER performance achieved by FAT-DL with $L=50$ is even better than that achieved by the LAMP algorithm with $L=65$. The performance gain comes from the fact that FAT-DL well incorporates the prior information of the Bernoulli-Gaussian mixture distribution and effectively boosts the AWGN precision auto-tuning. Moreover, the proposed FAT-DL algorithm effectively decreases the computational complexity by dimension reduction. This is an appealing observation since a small number of training data is required for the deep learning network.

\begin{figure}[h]
\centering
\includegraphics [width=0.5\textwidth] {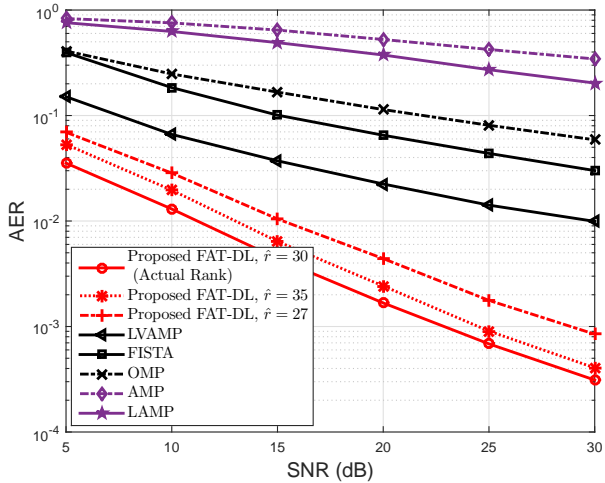}
\caption{The AER for different SNRs with $N = 100$, $\epsilon_{k}=0.3$, $M = 64$, $L=60$, and $\mathbf{A}$ has a mean $\mu = 7$.}
\label{snrmean1}
\end{figure}

Fig. \ref{snrmean1} plots the detection performance with different SNRs under the scenario of a mean-perturbed pilot matrix $\mathbf{A}$. It is observed that for the considered range of SNRs, the AMP algorithm, the FISTA algorithm, the OMP algorithm, the LAMP algorithm, and the LVAMP algorithm perform worse than the proposed FAT-DL algorithm, and the performance gap is enlarged as the SNR increases. The reason is that dimension reduction in \eqref{signalspace} becomes more accurate in high SNR region. Moreover, it is found that the proposed algorithm is not
sensitive to the accuracy of rank estimation when the estimated
rank is larger than the actual rank. This is because although
overestimating the rank leads to more noises to be included,
the transformed device state matrix based on the overestimated
rank contains the desired signal space of that based on the
actual rank. Compared with extra noises, the desired signal
space dominates the impact on the detection accuracy. The
performance gap between the proposed algorithm with actual
rank information and the one with overestimated rank is negligible for sufficiently long pilot sequences. Thus, we can utilize
a relatively large rank for guaranteeing the AER performance
if the rank estimation is less accurate. Similar to the LVAMP algorithm, the proposed FAT-DL algorithm is not sensitive to the mean value of the pilot matrix. However, the AER performance of AMP and LAMP degrades severely in the presence of a nonzero mean $\mathbf{A}$.

\begin{figure}[h]
\centering
\includegraphics [width=0.5\textwidth] {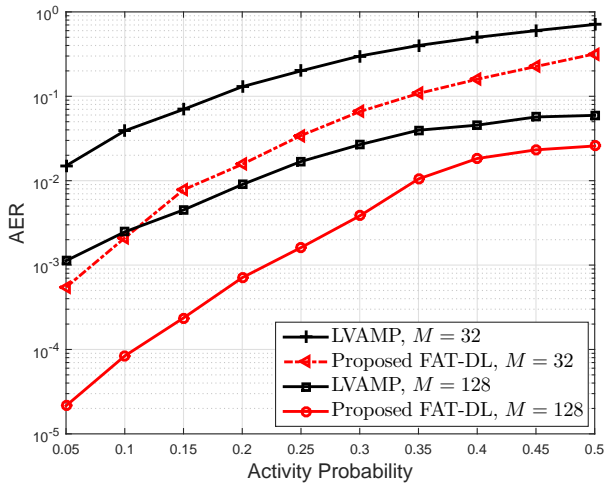}
\caption{The AER for different activity probabilities with $N = 100$, $L=60$, SNR = $15$ dB, and $\mathbf{A}$ is i.i.d. Gaussian distributed with zero mean and unit variance.}
\label{sparsityiid}
\end{figure}

Fig. \ref{sparsityiid} plots the AER curves of the considered algorithms against different activity probabilities under an i.i.d. Gaussian distribution $\mathbf{A}$ with zero mean and unit variance. It is seen that the performance of all the algorithms is degraded as the activity probability increases.
This is because the co-channel interference among devices increases as more devices are active. The proposed algorithm outperforms the LVAMP algorithm by a large margin even if the activity probability is higher than $0.45$. In practice, the proposed algorithm is appealing in various IoT applications with a wide range of activity probability. Fig. \ref{sparsityiid} also confirms that as the number of BS antennas, $M$, increases, the AER of the proposed FAT-DL algorithm drops rapidly and faster than that of the LVAMP algorithm, indicating that the proposed algorithm can quickly drive the detection error to zero with a fewer number of BS antennas for saving cost.

\begin{figure}[h]
\centering
\includegraphics [width=0.5\textwidth] {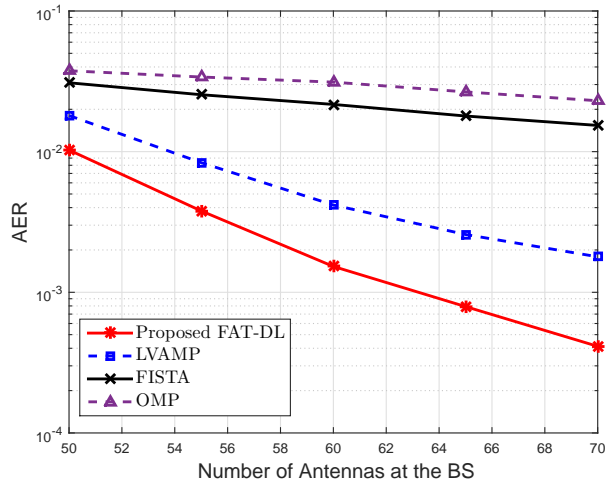}
\caption{The AER curve with respect to the number of antennas at the BS with $N = 1000$, $L=200$, $\epsilon_{k}=0.1$, SNR = $25$ dB, and $\mathbf{A}$ has a mean $\mu = 5$.}
\label{antenna}
\end{figure}

Fig. \ref{antenna} plots the error probability versus the number of antennas at the BS.
As expected, increasing the number of antennas at the BS improves the performance appreciably and the proposed FAT-DL algorithm achieves much better performance than that of the other considered algorithms. The performance gain stems from that the proposed algorithm not only exploits the prior information of the Bernoulli-Gaussian mixture distribution, but also boosts the precision auto-tuning to enhance the detection performance.
Actually, due to $\text{rank}(\mathbf{X})\leq \text{min}\{M,K\}$, when the BS is equipped with a large antenna array, the reduction of computational complexity of the proposed algorithm is substantial. Thereby, the proposed algorithm is appealing for massive MIMO regime, which is widely assumed in 6G wireless networks.

\begin{figure}[h]
\centering
\includegraphics [width=0.45\textwidth] {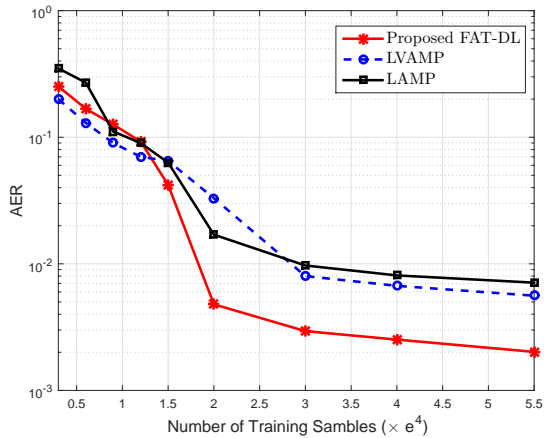}
\caption{The AER curve with respect to the number of training samples with $N = 100$, $\epsilon_{k}=0.1$, $M=32$, SNR = $15$ dB, $L=50$, and $\mathbf{A}$ is i.i.d. Gaussian distributed with zero mean and unit variance.}
\label{sample}
\end{figure}

Fig. \ref{sample} illustrates the impact of the number of training samples on the AER performance of the considered deep learning-based detection algorithms. In Fig. \ref{sample}, the size of the testing samples is the same as the size of the training samples. Initially, in the regime with a few numbers of training samples, the AER of the considered algorithms is poor and decreases sharply as the number of training samples increases. However, when the number of training samples continues to increase, the performance improvement diminishes. As a result, the curves are not smooth. The reason for this phenomenon is that for a small training set, samples cannot fully characterize the distribution features and overfitting occurs, which substantially increases randomness and degrades the AER performance. However, increasing the number of training samples helps decrease the AER and smooth the curves. Note that the saturation AER value of the proposed FAT-DL algorithm is lower than that of the other two algorithms. This is because the conventional algorithms attempt to detect the device only from the received signal, without exploiting prior observations. While the proposed FAT-DL algorithm well incorporates the prior information of the Bernoulli-Gaussian mixture distribution, and also effectively boosts the precision auto-tuning by combining the inner and outer networks. In addition, for the minimum required sample number of the proposed FAT-DL, LAMP, and LVAMP algorithm, $30,000$ samples are sufficient. Taking training efficiency, testing performance, and stability into consideration, we generate more samples for training, validation, and testing in other simulations.

\begin{figure}[h]
\centering
\includegraphics [width=0.45\textwidth] {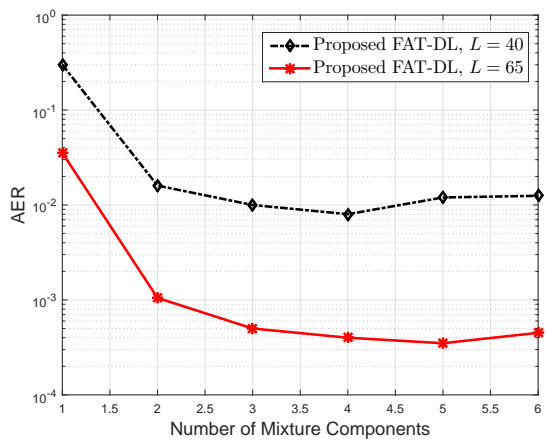}
\caption{The AER curve with respect to the number of mixture components with $N = 100$, $\epsilon_{k}=0.1$, $M = 32$, SNR = $15$ dB, and $\mathbf{A}$ has a mean $\mu = 5$.}
\label{component}
\end{figure}

In Fig. \ref{component}, we examine the sensitivity of the proposed FAT-DL algorithm
to the setting of the mixture component number $J$. We run the FAT-DL algorithm with $J$ varying from $1$ to $6$. It can be easily observed that when $J$ is $1$, underfitting occurs in the model which substantially degrades the AER performance. After $J$ is larger than $3$, the detection performance of the proposed algorithm
tends to be stable and is not very sensitive to the choice of the number of mixture components. Too big $J$ may cause overfitting effects in the model. Actually, in all our real experiments, we just simply set the mixture component number as $3$, and the proposed algorithm can consistently perform well throughout all our simulations.

\begin{figure}[h]
\centering
\includegraphics [width=0.5\textwidth] {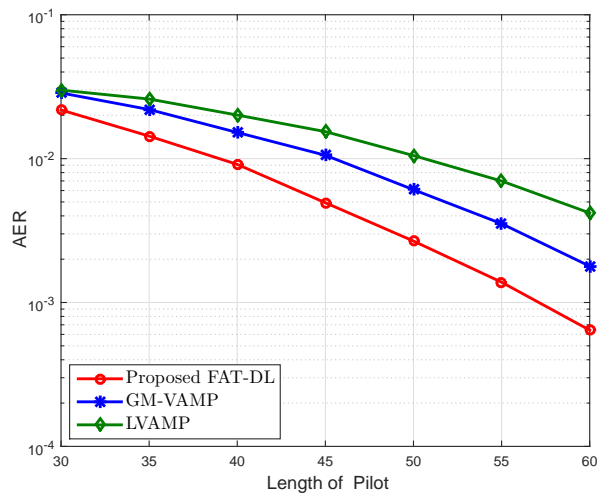}
\caption{The AER curve versus the length of pilot for exploiting effect of the fixed distribution parameters with $N = 100$, $M=64$, $\epsilon_{k}=0.1$, SNR = $20$ dB, and $\mathbf{A}$ has a mean $\mu = 5$.}
\label{constant}
\end{figure}

Next, we study the effect of the fixed distribution parameters on the AER performance of conventional algorithms, as compared to that of the FAT-DL algorithm. Specifically, we compare the FAT-DL algorithm with LVAMP and GM-VAMP algorithms in Fig. \ref{constant}. Herein, GM-VAMP refers to the VAMP algorithm proposed in \cite{vampr} where the soft-threshold denoiser is replaced by our designed Bernoulli-Gaussian mixture denoiser in \eqref{e6}, but the distribution parameters $\boldsymbol{\Omega}$ are set as fixed constants. Fig. \ref{constant} shows that the FAT-DL algorithm performs much better than  the LVAMP algorithm due to the fact that FAT-DL can approximate the distribution of device state matrix more accurately compared with the LVAMP algorithm utilizing the soft threshold denoiser. Importantly, it is observed that the AER of FAT-DL is smaller than that of the GM-VAMP algorithm. The performance gain stems from that the FAT-DL algorithm does parameter tuning based on sample space, while GM-VAMP employs the set of prior fixed parameters over all realizations, which are not exact any more in practical data. Such characteristics of the FAT-DL algorithm is mostly beneficial to combat the wrong distribution parameters effect.

\begin{figure}[h]
\centering
\includegraphics [width=0.5\textwidth] {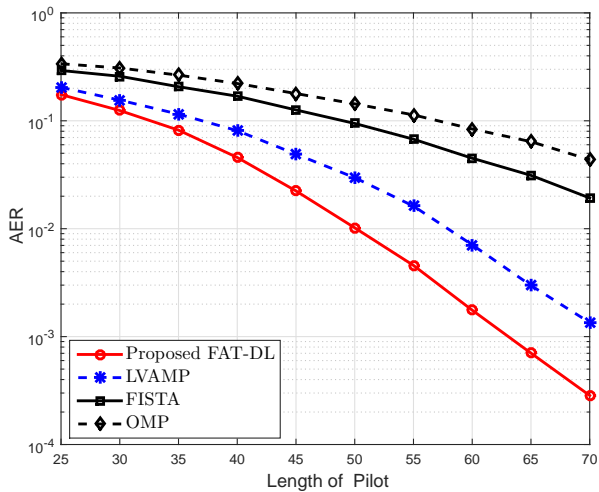}
\caption{The AER versus the length of pilot for recovery of Bernoulli-Student's-t signals with $N = 100$, $\epsilon_{k}=0.2$, $M = 32$, SNR = $20$ dB, and $\mathbf{A}$ is i.i.d. Gaussian distributed with zero mean and unit variance.}
\label{student}
\end{figure}

Finally, to show the AER for a different version of distribution of the samples, this paper postulates a Bernoulli-Student's-t distribution as a heavy-tailed prior on the
device state matrix, where the non-zero elements contain a few large values and many small ones. The probability density function (PDF) of the considered Bernoulli-Student's-t distribution is explicitly given by
\begin{align}\label{stu}
  &p(\mathbf{S})=  \prod_{n=1}^{N}\prod_{r=1}^{r^e}(1-\epsilon_{nr}  )\delta(s_{nr})\nonumber \\
  &+\epsilon_{nr} \frac{\Gamma((\nu +1)/2)}{\sqrt{\pi}\Gamma (\nu /2)}(1+s_{nr}^2)^{-(\nu+1)/2},
\end{align}
where $\Gamma(\cdot)$ is the Gamma function and the non-compressible rate $\nu$ is set as $1.9$. In Fig. \ref{student}, we observe that the OMP algorithm and the FISTA algorithm perform relatively poorly, the LVAMP algorithm performs relatively well, and the FAT-DL algorithm outperforms all other algorithms under the heavy-tailed distribution. We attribute this behavior to the FAT-DL algorithm's ability to tune itself and adapt to the signal at hand, namely the adopted denoiser based on Bernoulli-Gaussian mixture distribution can approximate the Bernoulli-Student's-t distribution well. Moreover, in the case of Bernoulli-Student's-t distribution, the FAT-DL algorithm has a smaller performance gain  over the LVAMP algorithm compared with the case in Fig. \ref{pilotmean1}. The reason is that the irregular distribution of samples leads to an increase in information loss. In other words, it may need more number of the mixture components in learning a more accurate approximation of the unknown parameters under Bernoulli-Student's-t distributed samples. Meanwhile, this comes with the expense of increased training time.

\section{Conclusion}
This paper has proposed a novel deep learning framework for massive device detection in $6$G wireless networks. The proposed framework contained a dimension reduction module, a deep learning network module, an active device detection module, and a channel estimation module. The dimension reduction module effectively decreased the computational complexity of massive device detection even with a large-scale antenna array at the BS. For the deep-learning network module, this paper designed a feature-aided adaptive-tuning deep learning network. Simulation results confirmed that the proposed algorithm can shorten the length of pilot sequences. Thus, the proposed deep learning framework is especially amenable to solve the high-dimensional device detection problem in $6$G wireless networks.

\end{document}